\documentstyle[11pt,paspconf,epsf]{article}

\begin{document}

\title{The Star Formation History in Nearby Dwarf Galaxies: the Fossil Record
in the Color-Magnitude Diagram}

\author{A. Aparicio}
\affil{Instituto de Astrof\'\i sica de Canarias, E38200 La Laguna,
Tenerife, Canary Islands, Spain}

\keywords{Dwarf Galaxies; Local Group; Star Formation History;
Color-Magnitude Diagram; Population Synthesis}

\begin{abstract}

\noindent Dwarf galaxies may play a key role in the formation and
evolution of bigger systems. This make a topic of major interest knowing how
they form and evolve and, in particular, how their star formation histories
(SFHs) have proceed since their birth. For nearby galaxies, the
color-magnitude diagram (CMD) contains stars formed over all their
lifetime. It is hence a fossil record of their SFHs. The analysis with
synthetic CMDs provides a powerful tool to retrieve them.

In this paper, I discuss the critical issues related to the computation
of synthetic CMDs, present a short summary of the currently
available results for the SFH extending the full life of galaxies and
make a few critical considerations about the powerfulness and back-draws
of the method.

\end{abstract}

\keywords{Star formation history; Local Group; Synthetic CMDs}

\section{Introduction}

At the low end of the luminosity function, dwarf galaxies seem to be by
far the most numerous in the Universe. At least, this is the case in the
Local Group, where 90\% of its about 40 members are dwarfs. Besides
this, the hierarchic scenarios of galaxy formation, such as an Universe
dominated by cold dark matter, predict that dwarf galaxies are likely to
be the first structures to form, arising from $1\sigma$ fluctuations in the
density distribution of the primeval Universe and that they would
preclude afterwards to form larger galaxies. These two reasons are
enough to pay a preferential attention to these objects. The fortunate fact that the
Local Group contains a wide variety of dwarf galaxies provides a good opportunity
to study them in depth and to test the predictions provided by Cosmology
about the formation and evolution of galaxies.

Containing stars born
over all the life-time of a galaxy, the color-magnitude diagram (CMD) is a
fossil record of the the star formation history (SFH), which is in turn
closely related with the evolution and, in last instance, with the
galaxy formation process itself. 
The best tool to deciphering that record is the analysis
with synthetic CMDs, which can simulate any arbitrary input SFH
(Aparicio et al. 1996; Gallart et al. 1999b). The fact that this work can
only be performed for the most nearby galaxies, makes
still more relevant  the study of the Local Group and its close
neighborhood.

In this paper, I will review the synthetic CMD method, its
virtues and its
shortcomings (Sec. 2 and 3); give a summary of results obtained for galaxies
for which the SFH extending their whole life is available through the
synthetic CMDs technique
(Sec. 4); make a concise statement about future work (Sec. 5)
and a short critical discussion of some of the
preconceptions and misconceptions related with
the SFH of galaxies and
its determination (Sec. 6).

\section{Star Formation History and Color-Magnitude Diagram}

\begin{figure}
\vspace{0cm}
\plottwo{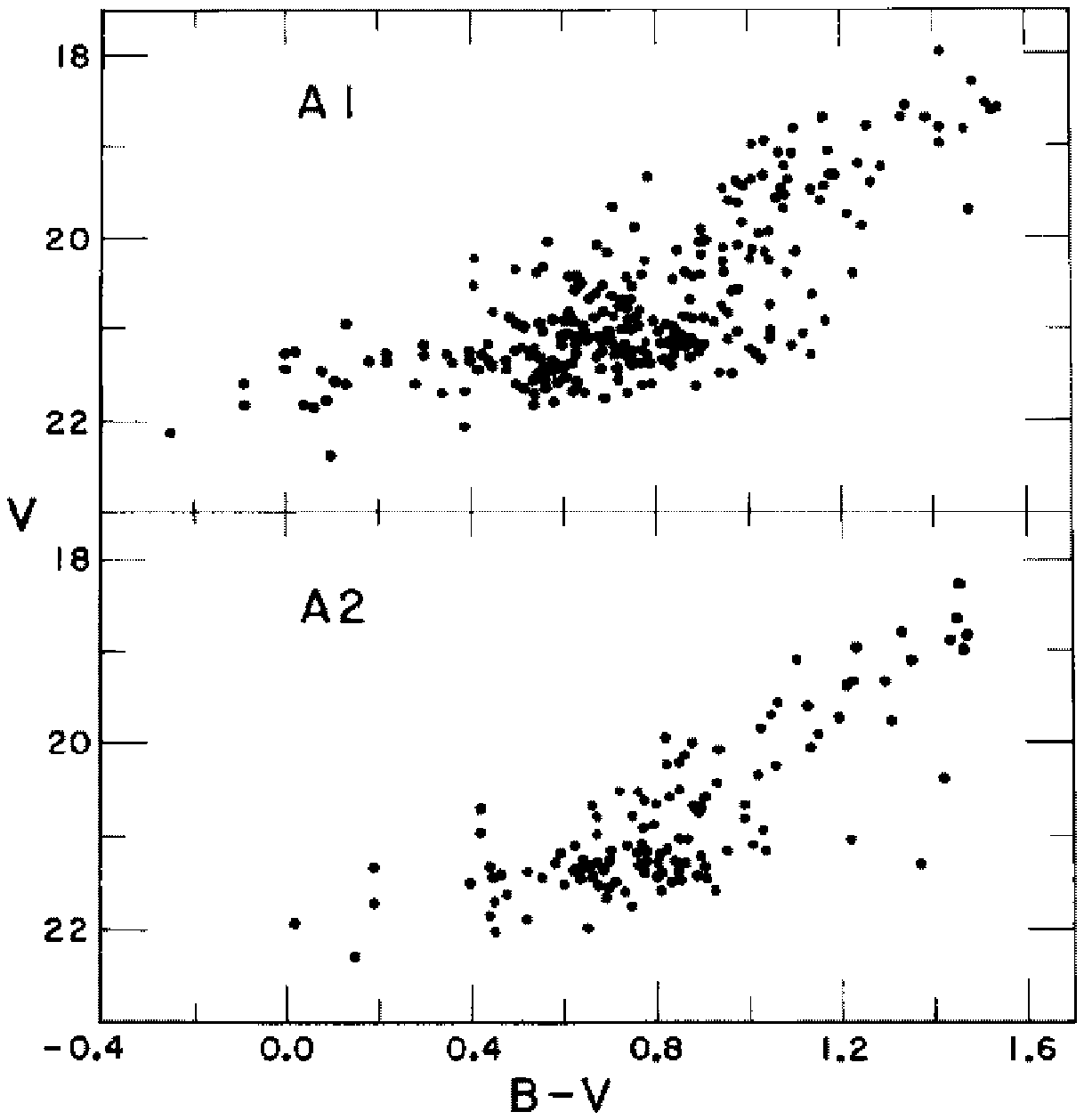}{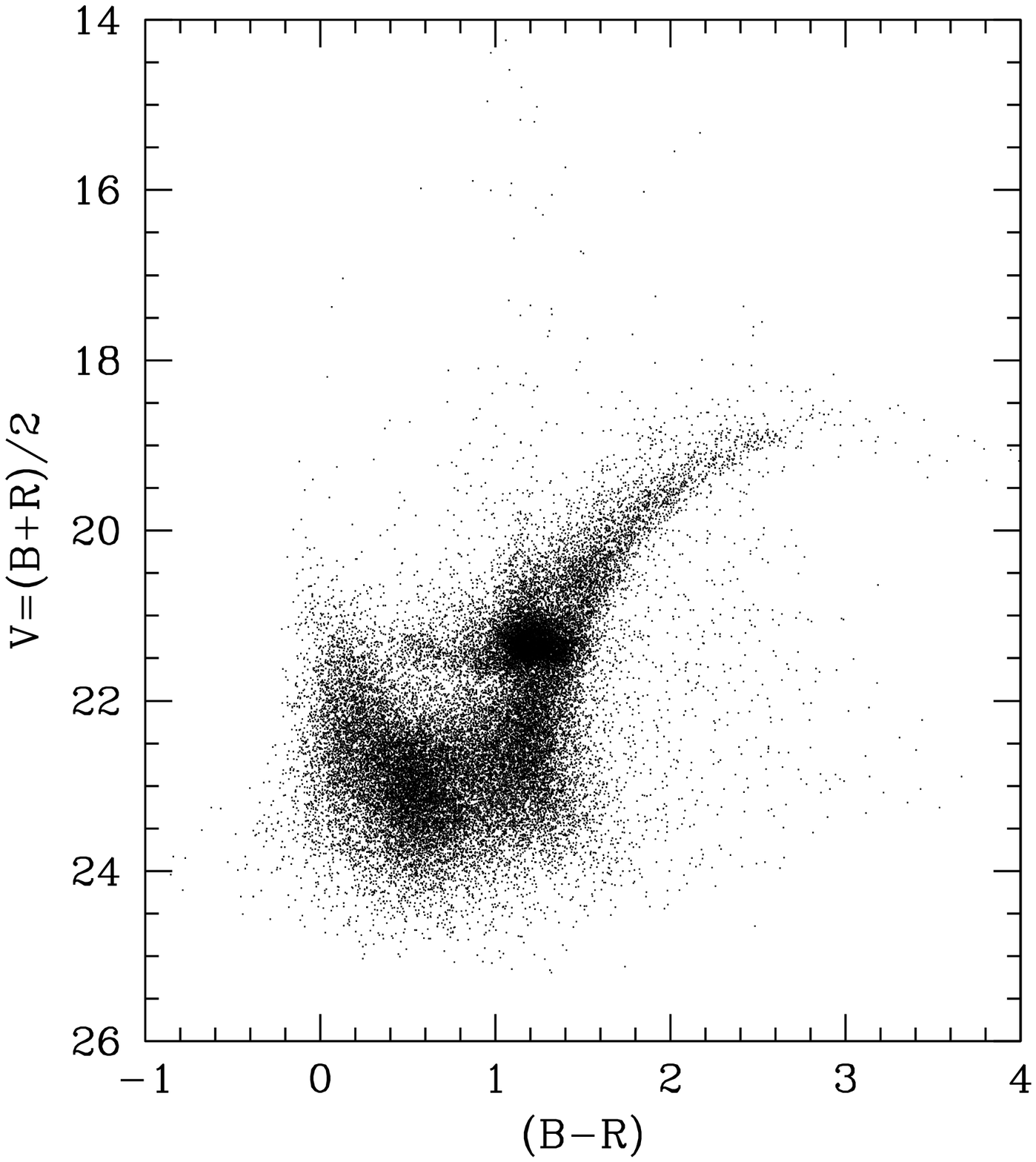}
\caption{CMDs of the Fornax dwarf galaxy showing the evolution in the
data quality in the last decade. Left panel is from
Buonanno et al. (1985); right panel is from Stetson et al. (1998). The
latter shows the present day state of the art for wide-field
ground-based observations.}
\end{figure}

\begin{figure}
\vspace{0cm}
\plottwo{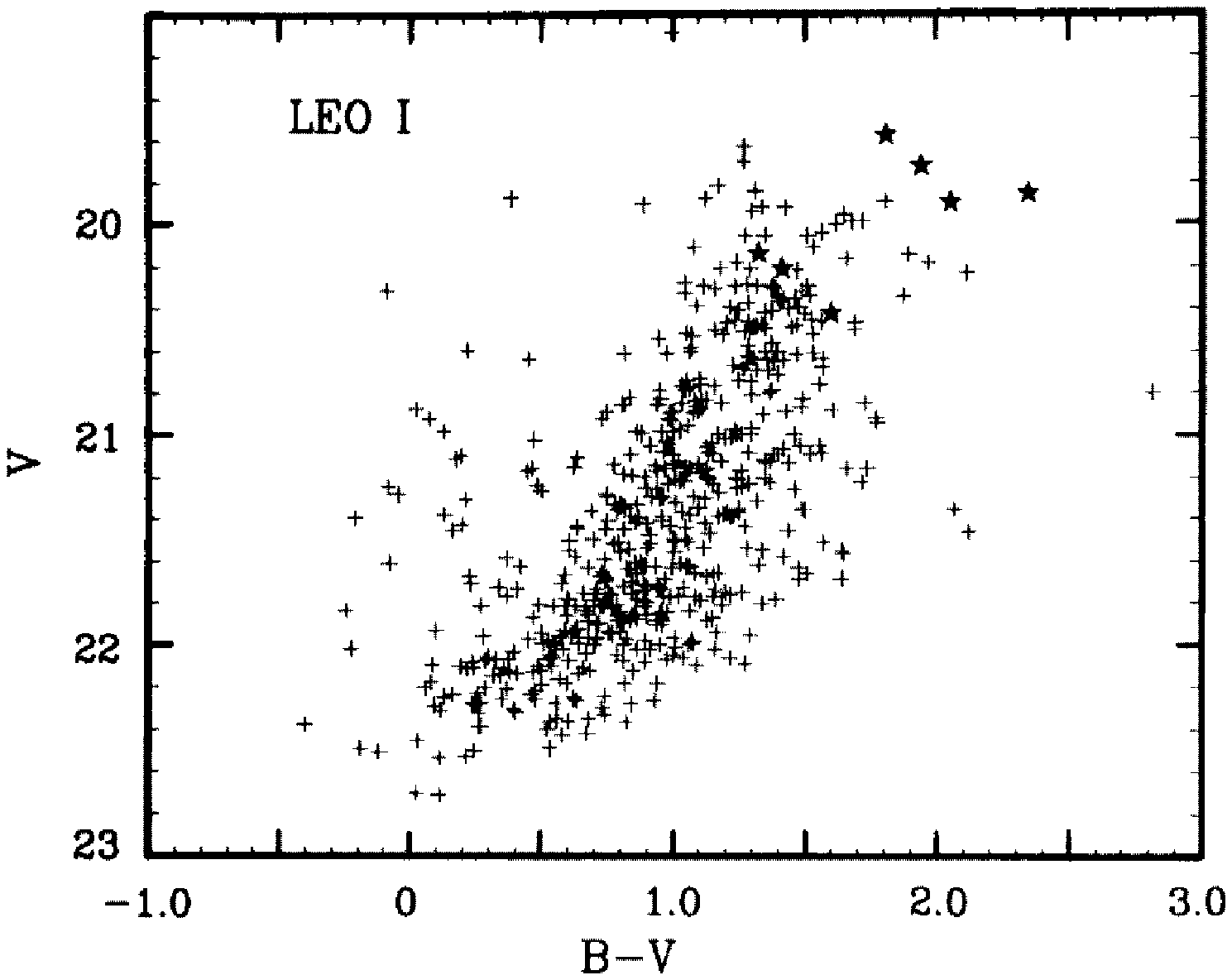}{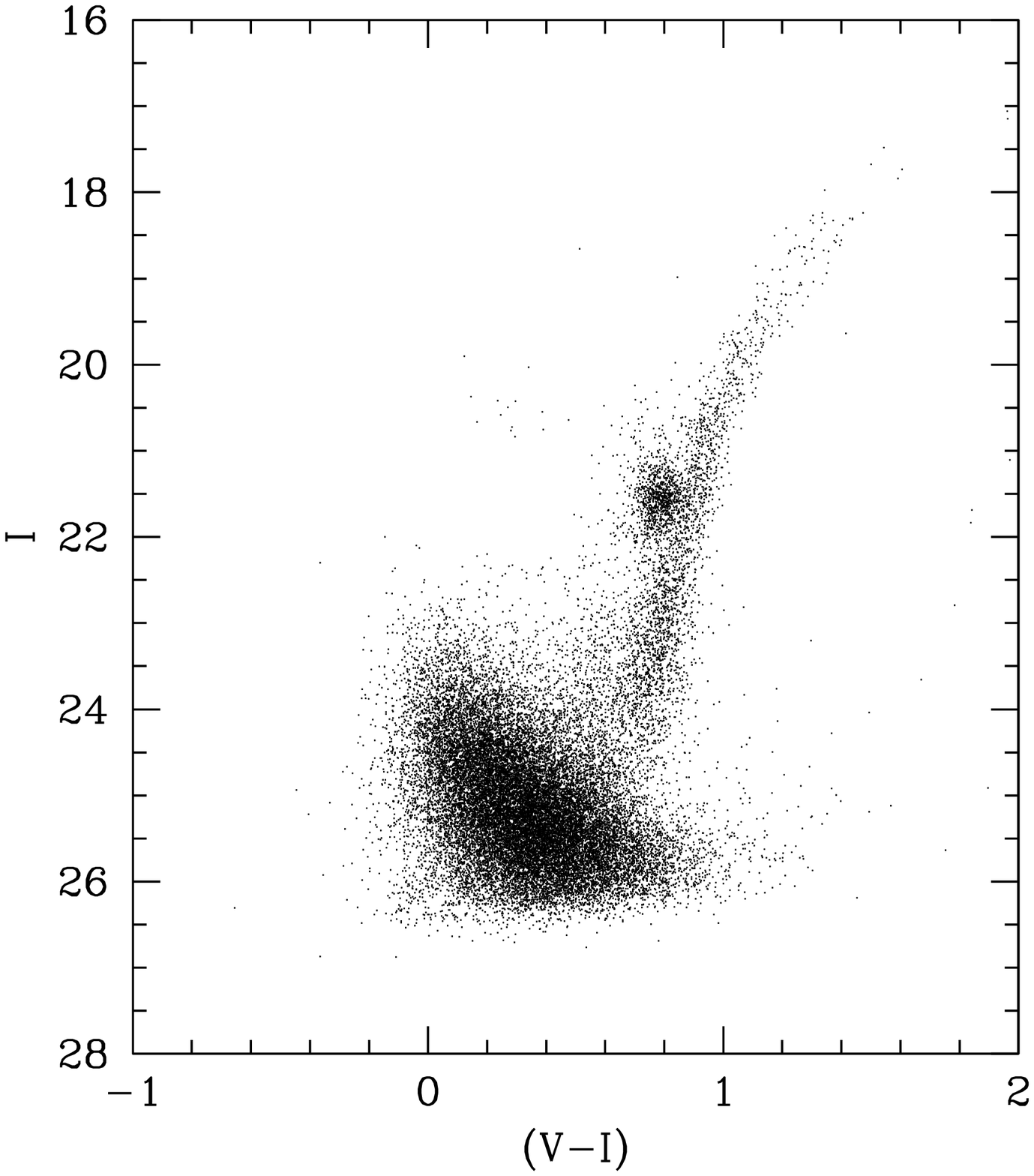}
\caption{CMDs of Leo I dwarf galaxy showing the evolution in the
data quality in the last decade. Left panel is from
Fox \& Pritchet (1987); right panel is from Gallart et al. (1999a). The
latter shows the present day state of the art for HST observations.}
\end{figure}

The recent growing interest on obtaining detailed SFHs of nearby galaxies has been, at least
partially, due to the fast development of our capabilities to determine them. This is linked to the
improvement of both our observational resources and our theoretical tools:

\begin{itemize}

\item From the observational side, the better image quality, instrument
sensitivity and field of view now available, compared with those of a decade ago, has
produced a great increase in the depth and accuracy of CMDs. To illustrate
this, Figs. 1 and 2 show the CMDs of two nearby dwarfs:
Fornax and Leo I. In both cases, left-side panels show a decade old data and
right-side panels, the latest available data. Not only the new data provide
information for the main sequence (MS) reaching the oldest stars, but also a
good sampling of fast stellar evolutionary phases, such as the
red-clump (RC) and blue-loop (BL) of core He-burning intermediate-age
stars, or the asymptotic giant branch (AGB). 

\item From the theoretical side, the improvement concerns two main issues:
(i) the availability of stellar evolution libraries with a wide coverage
in age and metallicity, and including the calculation of advanced
phases such as the horizontal-branch (HB) or the AGB and (ii) the
development of codes for computation of synthetic CMDs. Details
on the analysis with synthetic CMDs are given
in Sec. 3.

\end{itemize}

The information on the SFH contained in CMDs and the potentiality
of the synthetic CMDs technique is visualized in Figure 3. It
shows the stellar population of a simulated galaxy in which the star
formation had proceeded at a constant rate from 15 Gyr ago to date and
in which the metallicity had increased linearly from $Z=0.0001$ to
$Z=0.004$. In each panel, stars with ages in the given intervals (in Gyr) are plotted. The changes not only in the MS extension, but
also in the shape of other features like the RC, the HB, the BL, the red giant branch (RGB) or the AGB
are clear. It is the
analysis of the distribution of the stars across the CMD, which is related to these changes, that can be
used to derive the
SFH. For details about the features of the CMD and the age
distribution of stars populating it see Chiosi et al. (1992);
Aparicio, \& Gallart (1994); Aparicio et al. (1996); Da Costa (1998) and
Mateo (1998).

\begin{figure}
\vspace{0cm}
\plotone{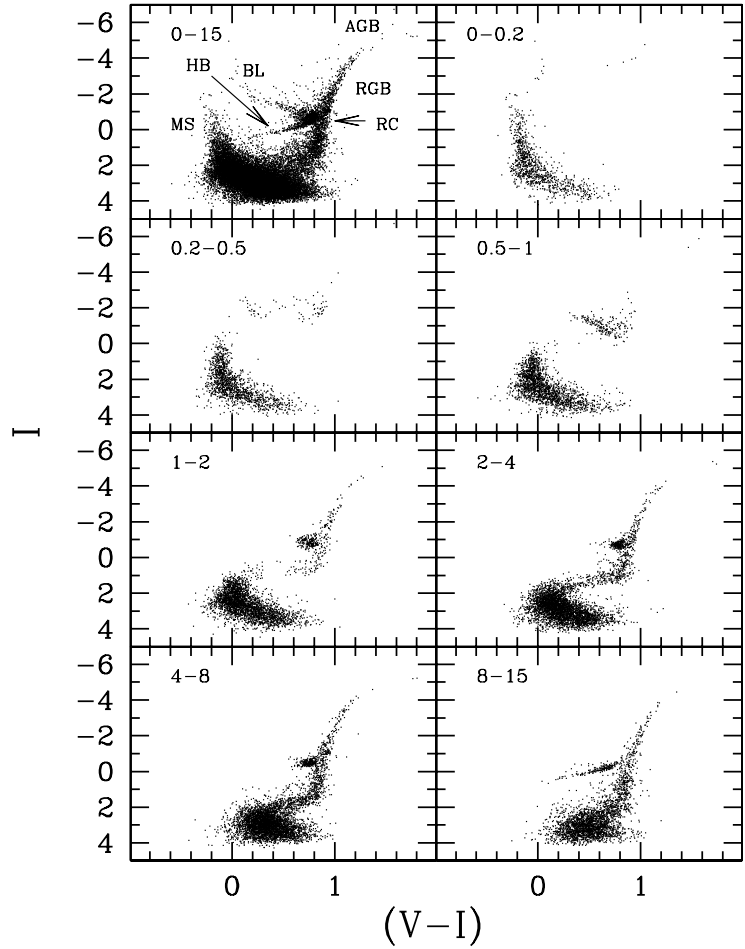}
\caption{Model CMDs for a constant SFR from 15 Gyr ago to the present time
and a linearly increasing $Z(t)$ from $Z=0.0001$ to
$Z=0.004$. In each panel stars with ages in the given interval (in Gyr)
are plotted. Observational effects have been simulated as described in
Sec. 3.2.}
\end{figure}

\section{Synthetic CMDs: the method}

The method to derive the SFH from synthetic CMDs is
composed of three main steps: (i) computation of the synthetic CMD
itself, reproducing a given SFH and using stellar evolutionary models;
(ii) simulation of observational effects into the synthetic CMD
and (iii) comparison of the observational and the synthetic CMD
to determine the acceptable and/or best models. Careful handling of the
problems arising in each of these steps is determinant in the success of
the final solution. I will shortly discuss these problems in the present
section.

\subsection{Computing the synthetic CMD}

The main input for the synthetic CMD computation is the SFH,
which can be understood as a combination of several simpler functions:
the SFR $\psi(t)$; the chemical enrichment law
$Z(t)$; and the initial mass function (IMF)
$\phi(m)$. Together with them, a function $\beta(f,q)$, controlling the 
fraction and mass ratio distribution of binary stars is also relevant. In the standard
procedure, a monte-carlo generator is then used to produce, for every star $i$, random
values of the mass $m_i$ and the time of birth $t_i$, according to $\phi(m)$ and $\psi(t)$, respectively.
Then, $t_i$ is used to determine the metallicity $Z_i$ through $Z(t)$. Later on, the binary nature of the
star and, eventually, the mass of its companion, is determined using also a
monte-carlo generator and $\beta(f,q)$. 

At this point, the stellar evolution models are used first to determine
whether the star is now alive. If so, an interpolation in the
age-mass-metallicity evolution grid space is performed to determine
first, the stellar evolution phase and then,
the luminosity and effective temperature of the star. Finally,
bolometric corrections are applied to obtain the magnitudes and color
indices. 

The critical point in this procedure is the interpolation inside a complete grid of stellar evolution models.
To achieve a good representation of the observed data, such interpolation must be done both in age and
metallicity to produce a smooth CMD containing the information of stars of any age and Z, and to be able to test any
desired chemical enrichment law. Otherwise, a lumpy synthetic CMD would result. It would not give a
good matching to the observational data and hence a reliable solution for the SFH could not be obtained.
Besides this, it is not superfluous to note that the stellar evolutionary models must include accurate
calculation of the shorter living phases, that can provide useful insight on the SFH. The RC, BL and
AGB are typical examples of such phases. 

The BL and the AGB deserve some comment. Both are quite sensitive to age
and metallicity and notably contribute to the age-metallicity resolution
when the lower MS is not accessible. However, colors of the BL stars
depend on metallicity through parameters difficult to control. As a
result, their color distribution can not be used as a confident
indicator of the absolute metallicity distribution, although their
luminosity distribution and their total number are good indicators of
$\psi(t)$ for stars younger than $\sim 1$ Gyr. Similarly, colors and
life-times of AGBs depend on the mass loss, again difficult to
parameterize, and on the bolometric corrections, ill-known 
for the very red stars. Nevertheless, their number and
relative distribution provide useful constraints on $\psi(t)$ for
intermediate to old ages. In summary, the BL and the AGB provide
good insight on the SFH, but uncertainties in the way they are
distributed in the CMD imply that only rough estimates can be obtained.

\subsection{Observational effects}

Mart\'\i nez-Delgado \& Aparicio (1997) show how dramatic the
consequences can be of 
neglecting observational effects in the analysis of a stellar population
with CMDs. Including all the sources of uncertainty affecting the photometry of
stars, observational effects must be simulated into the synthetic CMDs before comparison
with the observations. While some of these effects are produced at the detector level (random noise, flat-field, bad
pixels) or are simple results of statistics (poissonian signal to noise ratio), other, like
crowding, are intrinsic of the observed object (surface density, luminosity, and color distributions of
stars) or depend on specific observational conditions (diffraction limit for
space-based telescopes or the atmospheric seeing for ground-based telescopes). In fact, crowding is the
effect usually imposing the ultimate limit to the photometry.

Observational effects are mainly of three kinds:

\begin{itemize}
\item Loss of stars, which mainly affects faint objects. The fraction of
measured to existing stars as a function of magnitude is the
completeness factor $\Lambda$.
\item Magnitude and color shifts ($\delta$), which depend on the magnitude and
color of the stars. In general, stars are measured brighter
than they are; blue stars are measured redder and red stars are measured
bluer. 
\item Total (external) errors ($\Sigma$) of the photometry, which are larger than
the internal errors ($\sigma$) and which are not related to them in a simple
way. 
\end{itemize}

It has been extensively written about the artificial stars procedure used to characterize these effects
(Stetson 1994) and their trends and way to introduce them in the synthetic CMDs (Aparicio \& Gallart
1995; Gallart et al 1996a). 
Suffice to note here that $\Lambda$, $\Sigma$ and $\delta$ are related with
the distribution of color and magnitudes of the stellar population and with crowding, hence with the
stellar surface density and instrumental resolution. The adequate simulation of observational effects
requires the information of a large number of artificial stars, because $\delta$ and $\Sigma$ follow
bidimensional distributions that vary across the CMD. Figures 7 and 8 of Aparicio
\& Gallart (1995) show the trends of the observational effects on CMDs of artificial stars.

Provided that a large number of trials (typically several ten thousand
artificial stars) has been done, we have proposed a simple,
non-parametric method to simulate the observational effects in the
synthetic CMDs. It is described in detail in Aparicio \& Gallart (1995)
and in Gallart et al. (1996a). In short, the method consist in applying
to each star in the synthetic CMD the shifts in color and magnitude
observed in an artificial star picked-up randomly from a subsample of
artificial stars with similar color and magnitude. The powerfulness of
the method resides in the empirical approach, with no modeling or
analytical approximations introduced in the simulation process. It
provides a full representation of all the observational effects in the
synthetic CMD. 

A refinement of the method consist
in using the synthetic CMD itself as the list of artificial stars
to be injected in the real images. The advantage of this approach is
that it is free of sampling
distortions that would appear in the most densely populated regions of
the CMD (see Mart\'\i nez-Delgado et al. 1999).

\subsection{Solving the SFH: comparison of synthetic and observational
CMDs}

The final step is searching the SFH producing the model CMD that best
matches the observational one. Usually, the solution is not unique,
as a result of the many degeneracies intrinsic to the CMDs
themselves and uncertainties in the stellar evolution models. A strong
limitation comes from the age-metallicity degeneracy in the RGB and
can only be solved in a fully satisfactory way if accurate photometry of
the MS is available down to the oldest turn-offs. But besides these unavoidable
intrinsic limitations, the method followed to determine
which are the acceptable synthetic CMDs and/or best
reproducing the data
introduces a new source of uncertainty. 

Several of the methods used to compare model and observational CMDs
treat o consider the diagrams as bidimensional (or n-dimensional in the general
case if several color indices are used) distributions of points with
gaussian errors and use different techniques to search for good
models. Tolstoy \& Saha (1996) use bayesian inference to calculate the
likelihood of model CMDs to be good representations of the data (see
also Tolstoy et al. 1998 for a recent application); Serra-Ricart et
al. (1996) apply a neural network technique to minimize a $\chi^2$; Ng
(1998) propose a $\chi^2$ combined with a Poisson merit function to
search for the best model and Hern\'andez et al. (1999a) apply a
variational method to maximize the likelihood.

The advantage of these approaches is that they are general methods
requiring no human intervention for the comparison process: 
just two distributions of data points are compared with no more
assumptions. The back-draw, in my opinion, is that, as a matter of fact,
not all the data points can be considered in an equal base, because
different stellar evolutionary phases are known with different
accuracy. For
example, the color distribution and blue extension of BL of intermediate
mass and massive core He-burning stars, is quite sensitive to
metallicity and to not well known details in the stellar evolution
models. But the lifetime of the stars in this phase, which is a function
of mass and hence of age, is known with good confidence. If a method just
comparing distributions of points is used, a big weight is given to the
badly known details of the stellar evolution models and, consequently,
the resulting SFH would be affected by this. But if, due to the
uncertainties, these stellar evolutionary phases are removed from the
comparison, the information provided by the distribution
of these stars in luminosity is lost (see the discussion by Ng 1998).

It is for these reasons why we usually prefer methods of comparison
that allow human intervention to decide how different groups of stars
have to be considered. The draw-back is that results could contain
some subjectivity effects. The advantage is that suitable weights can
be given to each evolutionary phase. In our approach, a good fit of the position and shape of the main features in the CMD is also used to impose a
first limit to the range of possible solutions for the SFH. Then several
regions are defined in  the CMDs and the numbers of stars in them are 
compared through a $\chi^2$ or a least squares method. 
The key point of the approach is that the size and
location of the bins is decided considering the information provided by
the different stellar evolutionary phases. While a fine grain
binning is done in the MS area, larger boxes are used for the upper AGBs
and a set of bins in luminosity but with no resolution in color is
employed for the BL. A further advantage is that many age intervals can
be studied through stars in different stellar evolutionary phases. In
those cases, the solutions can be compared as a test of consistency.
Examples of studies of our group using this approach are Gallart et
al. (1996b,c, 1999b); Aparicio et al. (1997a,b); Mart\'\i nez-Delgado et al
(1999). Hurley-Keller et al. (1998) use a similar technique. A
conceptually analogous method was introduced by Bertelli et
al. (1992) (see also Vallenari et al. 1996). A binning procedure is also proposed by Dolphin (1997). An
early simplified approach, using only luminosity functions was used by the group of Bolonia (e.g. Tosi et
al. 1991). 

The former discussion deals with the method of comparison of 
CMDs itself, but nothing has been said about the SFH search process. In
many of the works that study real data, a set of model CMDs
is computed for several different input SFHs among which the best solution
is searched (Tolstoy 1996; Gallart et al. 1996a,b; Aparicio et
al. 1997a; Hurley-Keller et al. 1998). This approach is a way to
overcome the difficulty imposed by the many degrees of freedom of the 
problem but introduces an intrinsic limitation: the best solution may
not be among the checked SFHs. A way to compensate this
limitation is checking many models, selecting all those producing an
acceptable matching of the observational CMD and taking the average of
them as the solution. 

However, methods that do not require explicitly
computation of the model CMDs are preferable. They have the main formal
advantage that require very few or none initial assumptions about the
SFH. Examples of them applied to the analysis of real data are Dohm-Plamer et
al. (1997); Aparicio et al. (1997b); Mighell (1997); 
Gallart et al. (1999b) and Hern\'andez et al. (1999b). Dolphin (1997) shows a consistency test using only synthetic
data.

In our recent applications (Aparicio et al. 1997b; Gallart et al. 1999b;
Gallart et al., this book),
observational effects are simulated in the way described in Sec. 3.2,
and $Z(t)$, $\phi(m)$ and $\beta(f,q)$ are tested together with
$\psi(t)$. Only a
model CMD with constant $\psi(t)$ for the whole time interval is
required for each combination of $Z(t)$, $\phi(m)$ and $\beta(f,q)$. The fact
that the model CMDs keep a record of the age of every star is used. The distribution
of stars in the boxes defined in the model CMDs
are computed separately for narrow age intervals. For formal purposes
only, we will call $\Delta_i(t)$ the step functions defining these time
intervals: $\Delta_i(t)=1$ for $t$ inside the $i$-th time
interval; $\Delta_i(t)=0$ otherwise. 

The distribution of stars that would correspond to any $\psi(t)$ can be
calculated as a linear
combination of the distributions for every age interval:

\begin{equation}
N_j^m=A\sum_i\alpha_iN_{ji}^m
\end{equation}
\noindent where the subindices $i$ and $j$ refer to the narrow age
intervals and to the bidimensional bins defined in the CMD
respectively and the superindex $m$ stands for the distribution in the
model CMD. It is this global distribution which is compared with
the results for the observational CMD through

\begin{equation}
\chi^2=\sum_j\frac{(N_j^o-N_j^m)^2}{N_j^o}
\end{equation}

\noindent where index $o$ refers to the observational CMD. The SFR can be obtained as

\begin{equation}
\psi(t)=A\sum_i\alpha_i\psi_0\Delta_i(t)
\end{equation}

\noindent where $\psi_0$ is the constant SFR of the synthetic CMD we are
using. $\chi^2$ is
a multidimensional function depending on the n
$\alpha_{\rm i}$ coefficients defining $\psi(t)$. Minimizing $\chi^2$
gives the best $\psi(t)$ for the set of $Z(t)$, $\phi(m)$ and
$\beta(f,q)$ used.

\section{Application of the method: summary of galaxies with a
quantitative derivation of its full-life SFH}

A major goal of the analysis of the SFH of nearby galaxies is to compare
the properties and evolutionary scenarios for a wide variety of them. This
requires obtaining quantitative SFHs for the full life of as many
galaxies as possible. Figure 4 summarizes the results of this kind presently
available. It does not include some detailed works in which,
nevertheless, the full history of the object has not been
covered. Intentionally, the Magellanic Clouds have not been included
because they probably deserve specific reviews (see Zaritsky, this book
and Geha et al. 1998). The most comprehensive review of SFHs for
galaxies in the Local Group, including many additional 
qualitative estimates is by Mateo (1998). 

\begin{figure}
\vspace{-2cm}
\plotone{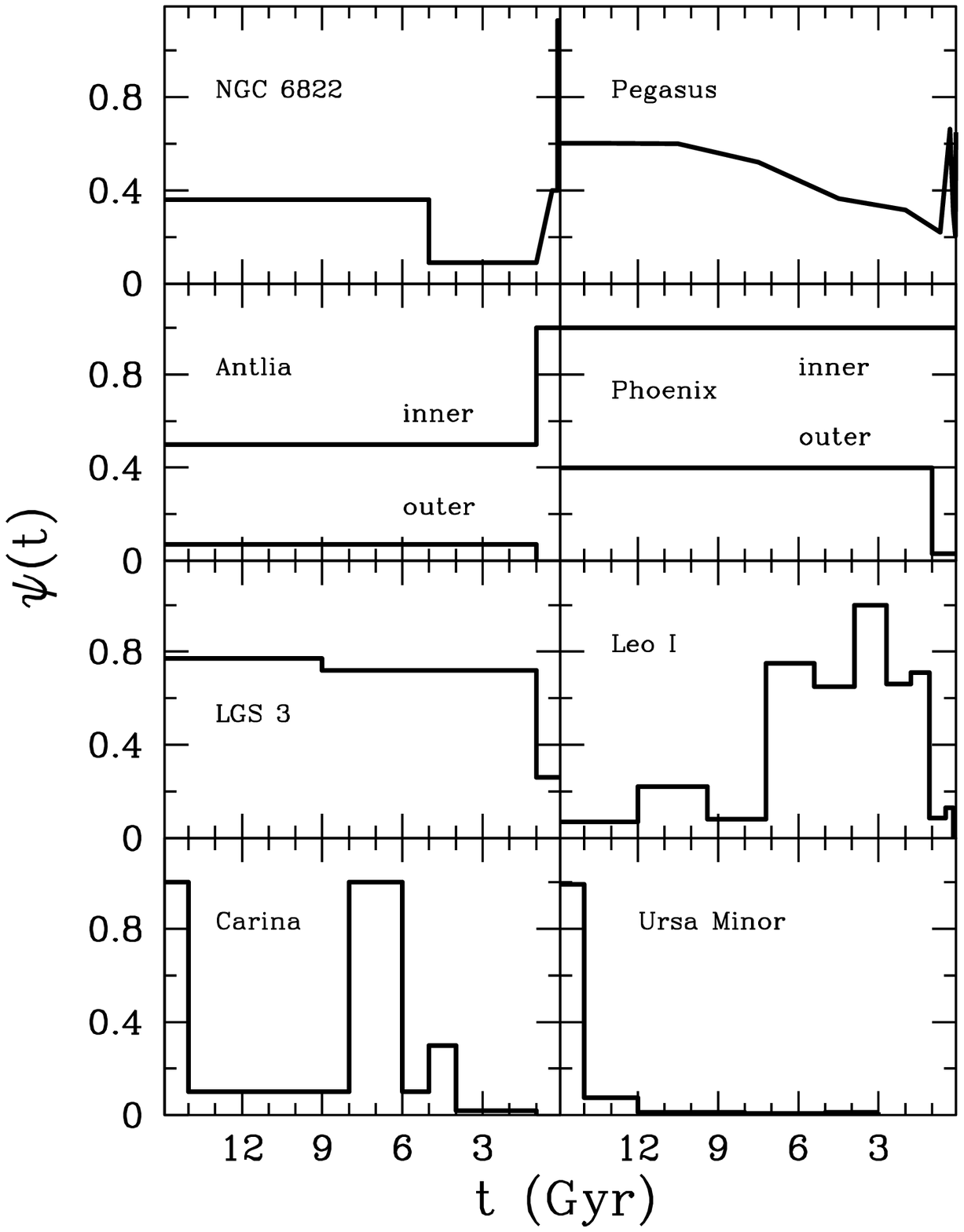}
\caption{Summary of SFHs of dwarf galaxies quantitatively determined and
extending the whole life of the system as presently available in the
literature. Normalization (M$_\odot$yr$^{-1}$pc$^{-2}$ that correspond
to the unity in the vertical scale), references and comments for each
galaxy follow:\protect\newline
\protect\newline
{\bf NGC 6822:} $12\times 10^{-9}$; Gallart et al. 1996b,c\protect\newline
{\bf Pegasus:} $3\times 10^{-9}$; Aparicio et al. 1997a. See also
Gallagher et al. 1998, specially for the youngest SFH\protect\newline
{\bf Antlia:} $6\times 10^{-10}$; Aparicio et al. 1997c\protect\newline
{\bf Phoenix:} $1\times 10^{-9}$; Mart\'\i nez-Delgado et al. 1999\protect\newline
{\bf LGS 3:} $2\times 10^{-10}$; Aparicio et al. 1997b\protect\newline
{\bf Leo I:} $2\times 10^{-9}$; Gallart et al. 1999b\protect\newline
{\bf Carina:} $5\times 10^{-9}$; Results from Mighell 1997
and Hurley-Keller et al. 1998 and structural parameters for Carina by
Mateo (1998) have been considered\protect\newline
{\bf Ursa Minor:} $2.5\times 10^{-9}$; Mart\'\i nez-Delgado et al., this book\protect\newline
}
\end{figure}

Galaxies in Fig. 4 are sorted by decreasing recent to
averaged SFR. A few general conclusions
can be drawn from it:

\begin{itemize}
\item Clear traces appear in all the galaxies of an important star
formation activity in the early epoch, regardless of whether the galaxy
is a dIr or a dSp/dE.
\item Dwarfs classified as dIr are characterized by the presence of an
enhanced star formation activity for the last few hundred Myr. This is
the case even for galaxies like Pegasus dIr or Antlia, that have been
sometimes considered as dIr/dSp transition objects.
\item While dSp/dE show no intense present day or very recent star
formation activity, the presence of a non-negligible intermediate-age to
young star formation activity is a common fact. Mechanisms must be found
for how these galaxies retain gas during extended periods of time to
produce further generations of stars.
\item For the young stars, where time resolution is much better,
bursting star formation seems to be present (see NGC 6822 and
Pegasus). This fact has also been beautifully illustrated for Sextans A
(Dohm-Palmer et al. 1997), GR8 (Dohm-Palmer et al. 1998)  and Leo A
(Tolstoy et al. 1998).
\item Evidence of relatively long periods of star forming activity
followed by long quiescent intervals is found in Carina (see also
Marconi et al. this book, for the case of Fornax).
\item In the cases where a wide field has been covered, gradients in the
stellar populations are found in the sense that the external regions of
the galaxies lack a recent star formation activity which is
present in the central region. This is the case of Antlia and Phoenix
and a similar result has been found in WLM by Minniti \& Zjilstra
(1996). Note however that a more accurate analysis of the stellar
population, including kinematics, is necessary prior to establish
whether the external population is a halo-like one, in the sense of
tracing the very early time of the galaxy's history.
\end{itemize}

\section{The next steps}

Among the several ways in which the research of the SFH can improve, I'd
like to emphasize the following three:
 
\subsection{Deeper}

HST can be used to obtain (i) deeper CMDs of the most nearby
objects reaching down to the MS turn-offs of the oldest stars and (ii) CMDs for galaxies in the range of
$\sim 5$ Mpc of similar depth of
those obtained from the ground for galaxies inside the Local
Group. Beautiful examples of both kind start now being
available. In particular, Gallart et al. (1999a,b) show for Leo I how
HST data can be used to derive reliable solutions
for the SFH (including $\psi(t)$, $Z(t)$, $\phi(m)$ and $\beta(f,q)$) of
the whole life of galaxies at distances up to several hundred Kpc. 
Hern\'andez et al. (1999b) use HST data of several Milky
Way satellites to obtain $\psi(t)$ from a general variational calculus
maximum likelihood method. Although these authors introduce
simplifications that are probably too strong (no chemical
enrichment; no binary stars; gaussian errors; most observational effects
are neglected), the method is a promising one and its full capabilities
are still to be developed. Other beautiful examples of detailed SFH,
but referred only to the last few hundred Myr are by Dohm-Plamer et
al. (1997); Gallagher et
al. (1998) and Tolstoy et al. (1998), among others.

On the other side, Lynds et al (1998) and Schulte-Ladbeck et al. 
(1998) show how HST CMDs of a galaxy like UGC 6456,
at a distance of $\sim 4.5$ Mpc
can be used to solve their SFHs with the same level of
reliability obtained from the ground for galaxies in the Local Group,
such as NGC 6822 (Gallart et al. 1996b,c) or Pegasus (Aparicio et al. 1997a). This is of great importance
because it shows the
feasibility of obtaining good estimates of the SFH 
of a large number of galaxies in the vicinity of the Local Group,
increasing the knowledge of the properties of galaxies in regions of
different densities.

\subsection{Wider field}

The new generation of wide field cameras
for ground-based telescopes can be exploited to study gradients and
differentiated disk-like and halo-like structures of the most nearby
galaxies. The analysis of Fornax by Stetson et al. (1998) and Phoenix 
by Mart\'\i nez-Delgado
et al. (1999) are first examples in this direction. Kinematic studies
carried on with 10 m class telescopes should provide fundamental
complementary information on the formation and evolutionary history of
these systems. 

\subsection{Physical scenarios for the evolution of galaxies}

From HST data, the SFR $\psi(t)$ and an estimate of the chemical
enrichment law $Z(t)$ can be reliably drawn
for the full life of the very nearby galaxies. Both functions can
be used together to provide clues on the physical evolution of the
galaxies in terms of their interaction with the intergalactic medium. 

The "real" or "true" yield $y$ is defined as the mass of newly-formed heavy
elements that a generation of stars ejects into the interstellar medium
per unit mass locked into stellar remnants or long-lived stars
($y=M_Z/M_\star$). If $y$ is known, the metallicity evolution of
the interstellar medium can be obtained through  (see Peimbert et al. 1994)

\begin{equation}
\frac{dZ}{dt}=\frac{ys\psi-Zf_{\rm I}}{M_{\rm g}}
\end{equation}

\noindent where $s$ is the fraction of mass that remains locked into
stellar remnants and long-lived stars in each generation of stars
(instantaneous recycling is assumed), $f_{\rm I}$ is the infall
rate and $M_{\rm g}$ is the gas mass. $Z$, $\psi$, $f_{\rm
I}$ and $M_{\rm g}$ are, in general functions of time. The gas mass can
be obtained as

\begin{equation}
\frac{dM_{\rm g}}{dt}=-s\psi+f_{\rm I}-f_{\rm O}
\end{equation}

\noindent with $f_{\rm O}$ being a function of time accounting for the
out-flow rate.

In summary, knowing $\psi(t)$ and $Z(t)$, even if they were just crude
estimates, and solving the former equations, limits can be put to
$f_{\rm I}$ and $f_{\rm O}$ and hence to the physical scenario of galaxy
formation and evolution through interaction with the intergalactic
medium. The necessary information to face these studies start now being
available and is of the kind of that provided by Gallart et al. (1999b)
for Leo I (see also Aparicio et al. 1997b). 

\section{Final considerations: old and new preconceptions and
misconceptions about the SFHs of dwarf galaxies}

To finish, I want to summarize a few topics that I consider
to be or to have been ill-established preconceptions or misconceptions
in relation to the SFH of dwarf galaxies. They are not
necessarily true or false. They are only concepts that I
see or have seen repeated here and there with no solid
justification. Some of them seem to be overcome by now. Others have been
discussed throughout this paper. Except for items
1 and 4 that are shortly developed in Sec. 6.1 and 6.2, I will only enunciate
them with no further comments.

\begin{enumerate}

\item Synthetic CMDs are a really very new thing (see Sec. 6.1).
\item Age-metallicity degeneracy largely avoids the possibility of
obtaining information about the SFH from the CMD.
\item Stellar evolution models are not good enough for deriving the SFH
from synthetic CMDs.
\item Blue plumes in dSp galaxies are certainly blue stragglers (see
Sec. 6.2).
\item dIr galaxies have only young stars.
\item dSp galaxies have only old stars.
\item Cosmology is important. The Local Group is not.

\end{enumerate}

\subsection{Concise prehistory of the synthetic CMD}

The following items show how old the idea of the synthetic CMD
is.

\begin{itemize}
\item The population concept of Baade (e.g. Baade 1944). In the opinion
of Sidney Van den Bergh (this meeting), this is probably the first step
towards the synthetic CMD analysis.
\item Maeder (1974) is in my knowledge the first to make and use a
synthetic CMD with a computational monte-carlo technique similar to the
one now in use.
\item Schild \& Maeder (1983) were the first in my knowledge to compute
monte-carlo synthetic CMDs applied to the study of the stellar
populations in galaxies.
\item The ZVAR era. ZVAR, by Bertelli and other members of the Padova
group is probably the code more extensively used for the purpose. Chiosi
et al. (1988), Aparicio et al. (1990) and Bertelli et al. (1992) are
representative examples of early use of ZVAR. The last is the first
example
of derivation of the full SFH of a nearby galaxy (the LMC) with
synthetic CMDs. It uses the two-dimensional distribution
of stars in the CMD and means an important step forward in the
accuracy with which the SFH of the LMC was known.
\item The Bolonia contribution marked a step forward in the systematic
study of the recent SFH of galaxies with monte-carlo synthetic CMDs. Tosi et al. (1991) is the first work
of this group on this topic.
\item And so on...
\end{itemize}

\subsection{Blue stragglers?}

About this topic, I just reproduce a sentence by Kenneth Mighell in the
coffee-line: "after time and work, every blue-straggler sequence in a
dSp has become a complex, intermediate-age population".

\vspace{4mm}
\acknowledgements

The expenses of my participation in this conference have been covered by
the organization of the meeting, the IAC (grant P3-94) and the DGICyT (grant
PB94-0433). I am also very grateful to the persons involved in the
organization of the meeting for their permanent and very useful help to
solve every kind of small and big problems. I am grateful to
Drs. Buonanno, Pritchet and Stetson for allowing me reproducing their
CMDs in Figs. 1 and 2.

\end{document}